\documentclass[nofigure]{pasj00}
\draft
\begin{document}
\SetRunningHead{S. Okamura et al.}{ICPNe in the Virgo Cluster}
\Received{2002/09/10}
\Accepted{2002/00/00}

\title{Candidates for Intracluster Planetary Nebulae\\ in the Virgo
Cluster based on the Suprime-Cam\\ Narrow-Band Imaging in {O[III]} and
H$\alpha$}

\author{%
   S. \textsc{Okamura}\altaffilmark{1,2},
   N. \textsc{Yasuda}\altaffilmark{3},
   M. \textsc{Arnaboldi}\altaffilmark{4,5},
   K. C. \textsc{Freeman}\altaffilmark{6}  \\


   H. \textsc{Ando}\altaffilmark{3},
   M. \textsc{Doi}\altaffilmark{7,2},
   H. \textsc{Furusawa}\altaffilmark{8},
   O. \textsc{Gerhard}\altaffilmark{9},
   M. \textsc{Hamabe}\altaffilmark{10},
   M. \textsc{Kimura}\altaffilmark{11,12},\\
   T. \textsc{Kajino}\altaffilmark{3},
   Y. \textsc{Komiyama}\altaffilmark{8},
   S. \textsc{Miyazaki}\altaffilmark{8},
   F. \textsc{Nakata}\altaffilmark{1},
   N. R. \textsc{Napolitano}\altaffilmark{4},\\
   M. \textsc{Ouchi}\altaffilmark{1},
   M. \textsc{Pannella}\altaffilmark{4},
   M. \textsc{Sekiguchi}\altaffilmark{11},
   K. \textsc{Shimasaku}\altaffilmark{1,2},
   M. \textsc{Yagi}\altaffilmark{3}
}%
\email{okamura@astron.s.u-tokyo.ac.jp}

\altaffiltext{1}{Department of Astronomy, School of Science,
University of Tokyo, Tokyo 113-0033, Japan}
\altaffiltext{2}{Research Center for the Early Universe, School of
Science, University of Tokyo, Tokyo 113-0033, Japan}
\altaffiltext{3}{National Astronomical Observatory of Japan, Mitaka,
Tokyo 181-8588, Japan}
\altaffiltext{4}{I.N.A.F., Osservatorio Astronomico di Capodimonte,
80131 Naples, Italy}
\altaffiltext{5}{I.N.A.F., Osservatorio Astronomico di Pino
Torinese, 10025 Pino Torinese,Italy}
\altaffiltext{6}{R.S.A.A., Mt. Stromlo Observatory, 2611 ACT, Australia}
\altaffiltext{7}{Institute of Astronomy, University of Tokyo, Mitaka,
Tokyo 181-0015, Japan}
\altaffiltext{8}{Subaru Telescope, National Astronomical Observatory
of Japan, Hilo, Hawaii 96720}
\altaffiltext{9}{Astronomisches Institut, Universit\"at Basel,
Venusstrasse 7, CH-4102 Binningen, Switzerland}
\altaffiltext{10}{Department of Mathematical and Physical Sciences,
Japan Women's University, Tokyo 112-8681, Japan}
\altaffiltext{11}{Institute for Cosmic Ray Research, University of
Tokyo, Kashiwa,Chiba 277-8582, Japan}
\altaffiltext{12}{present address: Department of Astronomy, Kyoto
University,
Sakyo-ku, Kyoto 606-8502, Japan}

\KeyWords{galaxies:clusters:individual (Virgo), intracluster planetary
nebulae,
cosmology: baryon fraction} 

\maketitle

\begin{abstract}
We have identified 38 candidates of intracluster planetary nebulae
(ICPNe) in a $34'\times27'$ field in the core of the Virgo cluster based on
the Suprime-Cam imaging through two narrow-band filters centered
at the redshifted wavelengths of the {[OIII]}$\lambda=5007\AA$
and the H$\alpha \lambda=6563\AA$ lines. Broad-band images in
$V$ and $R$ bands are used to check for any emissions in the adjacent
continuum. We describe the method briefly and present the list of
intracluster planetary nebulae candidates, together with their
finding charts. The ICPN candidates show a highly inhomogeneous
distribution,
which may suggest an association with the M86-M84 subcluster.
Fraction of diffuse intracluster light with respect to total light in
galaxies is estimated to be about $10\%$, leading to an estimate
of about 20\% for the baryon fraction. Spectroscopic follow up and a wider
survey are critical to reveal the nature of intracluster stellar population.
\end{abstract}

\section{Introduction}

Recent observations show that there is a substantial amount of
{\it intracluster stellar population}, which is observed as
diffuse intracluster light (\cite{Bernstein1995}), or as
individual stars, {\it i.e.,} planetary nebulae (\cite{Arnaboldi1996};
\cite{Theuns1997}; \cite{Feldmeier1998}),
and red giant stars (\cite{Ferguson1998}; see also the review
by \cite{Feldmeier2002})

The presence of this intracluster stellar population is
expected from dynamical processes which unbind stars from galaxies,
as predicted by interactions among galaxies ('harassment')
in a cluster potential (\cite{Moore1996}).
However, it is also possible that some stars were formed in more or
less uniform intracluster matter before the matter was assembled in
individual galaxies, or as the protogalaxies were falling in
the cluster potential (\cite{Merritt84}). Those stars, if present, would
have
the properties similar to those of the hypothetical population III.
Detection and a quantitative estimate of the amount of such a hypothetical
population would have a profound impact on our understanding of the
formation and evolution of galaxies and clusters of galaxies.\\
The presence of a diffuse stellar component in clusters is relevant
also for the ongoing discussion of the baryonic mass in clusters
(\cite{Fukugita98}) and the efficiency of star formation
(\cite{Balogh01}).
To account for the diffuse component, the total baryonic mass of a
galaxy cluster, $M_b$, should be written as:
\begin{equation}
M_b=M_{\rm gal} + M_{\rm gas} + M_{\rm IC*},
\end{equation}
where $M_{\rm gal}$ represents the mass of stars and interstellar
matter which reside in galaxies, $M_{\rm gas}$ is the mass of hot
intracluster gas, and $M_{\rm IC*}$ is the mass of intracluster stars.
So far, $M_{\rm IC*}$ was neglected, while recent studies
argue that it may be a significant fraction of the total baryonic
mass, at least 15\% and perhaps much more (\cite{Feldmeier1998};
\cite{Arnaboldi2002a}).

In order to address the origin and constraint the amount of the intracluster
stellar population, we need to establish the properties of
this population of stars.
Intracluster planetary nebulae (ICPNe) are excellent tracers of the
intracluster stellar population (\cite{Arnaboldi1996}; \cite{Freeman2000};
\cite{Arnaboldi2002a}). Searches for ICPNe were carried out so far using a
narrow band filter centered on the {[OIII]} $\lambda5007\AA$
line (\cite{Theuns1997},
\cite{Ciardullo1998}, \cite{Feldmeier1998}, \cite{Feldmeier2002},
\cite{Arnaboldi2002a}). However, the first spectroscopic follow-up
carried out on a subsample of ICPNe candidates selected via this
technique showed a significant amount of contamination
(\cite{Kudritzki2000}, \cite{Freeman2000}).
One class of contaminants consists of continuum objects
misclassified as line-emitters because of photometric errors.
The other class includes high redshift line emitters such as
{[OII]} starbursts at $z\sim0.347$ and Ly$\alpha$ emitters at
$z\sim3.13$.
Contamination by continuum objects can be solved by taking
an adequate off-band images, but reducing contamination from high-z
line emitters is a much more difficult problem.

To overcome the latter problem, we made a search for ICPNe in the
Virgo cluster using two narrow band filters centered on the
redshifted {[OIII]} line and H$\alpha$ line, the strongest and the
second strongest emission line respectively from a PN,
with the Suprime-Cam on the 8.2m Subaru
telescope. Such a challenging program can be carried out only with
an 8 meter telescope equipped with a wide-field imager
because H$\alpha$ emission from a PN is $3 - 5$ times weaker than
the strongest $\lambda$ 5007 [OIII] green line.

We present in this Letter the first results of our ongoing survey and
the list of secure candidates for ICPNe in the core of the Virgo cluster,
together with their finding charts. New estimates are provided for the
lower limit of the intracluster stellar population and the fraction
of baryonic matter in the Virgo cluster core.
Extensive discussion of the candidate selection criteria and
the results of the spectroscopic follow up are given in
\citet{Arnaboldi2002b}.

\section{Observation and Data Reduction}

In March-April 2001, a field in the central region of the Virgo
cluster was observed during the commissioning of the Suprime-Cam
10k$\times$8k mosaic CCD camera (Miyazaki et al. 2002) at the prime
focus of the 8.2m Subaru telescope. The camera covered an area of
$34'\times27'$ with a resolution of $0.2''$ per pixel.\footnote{one
2k$\times$4k chip was dead in the March run.} The field is
just south of M84-M86 ($\alpha=12^h25^m47.^s0$,
$\delta=+12^{\circ}43'58''$: J2000).

The field was imaged through two narrow band filters which
have ($\lambda_c$, $\Delta\lambda$)=(5021$\AA$, 74$\AA$) and (6607$\AA$,
101$\AA$),
corresponding to the redshifted {[OIII]} and H$\alpha$ lines, respectively.
Two standard broad band filters ($V$ and $R$) were also used to check the
intensity in the adjacent continuum.
Total exposure times are 900 sec, 720 sec, 3600 sec, and 8728 sec for
$V$, $R$, {[OIII]}, and H$\alpha$, respectively.
The seeing was slightly better for the narrow bands ($0.''6-0.''7$)
than for the broad bands ($0.''75-1.''0$).

Data reductions, including an astrometric solution, were carried out
with a data reduction package developed by the Suprime-Cam team.
All the reduced images in the same band were coadded and normalized
to 1 sec exposure. We then used a combined ($V+R$) continuum image,
following the procedure adopted by \citet{Steidel2000} to check for any
emission in the continuum.

We use SExtractor (\cite{Bertin1996}) to carry out the detection and
photometry.
Because the {[OIII]} image is the deepest, we perform the source detection
on the {[OIII]} image and then compute aperture magnitudes for the
H$\alpha$ and $(V+R)$ images at the location of the {[OIII]}-detected
objects.

Our selection criteria are based on our instrumental
magnitudes. Limiting magnitudes and detection limits are derived
from the simulation described in \citet{Arnaboldi2002a}, and shown
in Table 1. The former
is the magnitude at which 50\% of the input sample is retrieved from
the simulated image, while the latter is the magnitude at which the fraction
of
retrieved point sources becomes zero.

\begin{table}
  \caption{Limiting magnitudes and detection limits in the instrumental
  magnitudes}\label{maglim}
  \begin{center}
    \begin{tabular}{cccc}
     \hline
      Band & $(V+R)$ & {[OIII]}& H$\alpha$ \\
      \hline
      $m_{lim}$ &-2.3 & 1.8 & 1.2 \\
      $m_{det}$ &-2.0 & 2.1 & 1.5 \\
      \hline
    \end{tabular}
  \end{center}
\end{table}

The calibration of our $\lambda$ 5007 [OIII] fluxes to the
$m(5007)$ magnitude system by \citet{Jacoby1989} is addressed
in \citet{Arnaboldi2002b}. We report here that
$m(5007)= m({\rm [OIII]})+26.30\pm0.05$ and $V = V_{inst} + 28.0\pm0.05$,
where $m({\rm [OIII]})$ and $V_{inst}$ are the instrumental magnitude for
a 1 second exposure in the {[OIII]} and $V$ bands respectively.

\section{Selection Criteria for ICPNe Candidates}
We have developed a robust technique for the identification of ICPNe
based on the two-color diagram, ({[OIII]} - H$\alpha$) versus
({[OIII]} - (V+R)).
We can classify the objects we detected into four broad classes
according to their location on the two-color diagram;
(a) secure PNe candidates which are detected in both {[OIII]} and H$\alpha$,
(b) high excitation PNe or high EW Ly$\alpha$ which are
detected in {[OIII]} and have very weak (due to residuals in the sky
subtraction) or negative flux in both $(V+R)$ and H$\alpha$
(This class probably has a small amount of contamination by
high-z emission line galaxies),
(c) high-z galaxies with Ly$\alpha$ or {[OII]} emission in our {[OIII]}
band, with some flux in $(V+R)$ and weak or negative flux in H$\alpha$,
(d) continuum objects such as ordinary stars and galaxies.
Full discussion and quantitative criteria are given in
\citet{Arnaboldi2002b}.

Our field includes two bright Virgo ellipticals, M86 (NGC4406) and M84
(NGC4374), which were surveyed for PNe by \citet{Jacoby1990} using the
on-off band technique, {\it i.e.,} only one narrow band filter centered on
the
redshifted [OIII] line. We matched the list of published PNe
candidates with the sources that we detected. We classified the matched
objects and obtained the results summarized in Table 2.
\footnote{M86 is partially out of our image and we could match only about
1/3 of the Jacoby et al. (1990)'s sample.}

\begin{table}[h]
  \caption{Statistics of the sample matched with Jacoby et
al.(1990)'s.}\label{match}
  \begin{center}
    \begin{tabular}{ccccc}
     \hline
        Galaxy & Matched & \multicolumn{3}{c}{class} \\
        \cline{3-5}
        & total   & (a) & (b) & (c)+(d) \\
     \hline
     M84 & 74& 19& 26& 29 \\
     M86 & 64& 13& 15& 36 \\
     \hline
    \end{tabular}
  \end{center}
\end{table}
We note that the identification of PNe in individual
galaxies via the on-off band technique and the "by-eye" identification
produce catalogues with a significant fraction,
39\% and 56\% in M84 and M86 respectively, of continuum objects
and line emitters with detected continuum, which we know cannot be
true PNe.

Table 3 summarizes the secure candidates of ICPNe identified in this study,
{\it i.e.,} those which are classified as (a) and are located
outside of $4r_e\times4r_e$ areas centered on M84, M86 and NGC 4388,
where $r_e$ is the effective radius in which a half of the total
muminosity is included.
Over-luminous objects in the M84 fields are also included because they may
be ICPNe on the near side of the cluster, according
to \citet{Ciardullo1998}.\footnote{A large depth of the Virgo cluster along
the line of sight was pointed out by \citet{Yasuda1997}.}

\section{Projected Distribution of the Candidates}

Distribution the ICPN candidates projected on the sky is shown
in Figure 1 on our deep {[OIII]} image. The lower left area
bounded by the black line indicates the area excluded from
our survey because of the dead CCD chip.

Figure 1 shows a remarkably inhomogeneous distribution of the ICPN
candidates. Even when the over-luminous objects are not considered,
the overdensity in the upper right quadrant of our field is highly
significant. The majority of the candidates seem to be related with the
M86-M84 region of the Virgo cluster, which suggests a local origin
for the ICPNe. They may be bound in the halos of these galaxies,
or they may have become unbound from these galaxies while they are
falling into the Virgo cluster ({\it e.g.,} \cite{Binggeli1993};
\cite{Rangarajan1995}). If the latter is the case, the highly
inhomogeneous distribution is possibly a hint of a very recent
harassment event. There are also several candidates which do not
seem to be associated with bright galaxies.

Two of the three dominant galaxies have radial velocities largely
different from the mean velocity of the Virgo cluster
(1050$\pm$35 kms$^{-1}$; \cite{Binggeli1993}); M84 (1060 kms$^{-1}$),
M86 ($-244$), and NGC 4388 (2524).
A spectroscopic follow up and a wider survey are badly needed
to clarify the nature of the parent stellar population of ICPNe.

Figure 2 presents the finding charts of the 38 candidates.

\section{Fraction of Diffuse Light and Baryon Fraction}

Our goal is to estimate the fraction of diffuse light coming
from the intracluster stellar population and the baryonic fraction
in the Virgo cluster core, where we have our survey data.

The crucial parameter in this evaluation is the PN specific
frequency parameter $\alpha_{1.0}$ which gives the number of PNe
per unit stellar B luminosity
within the first 1.0 mag of the bright end cut off of the {[OIII]} PN
luminosity function, as selected according to our criteria.
There is no theoretical prediction for its value which may well
depend on the age/metallicity of the stellar population
from which the PNe originate. Its value has always been derived so far
from empirical measurements.
\citet{Arnaboldi2002b} estimated the parameter using 45 PNe in M84 as
$$\alpha_{1.0} = 3.46 \times 10^{-9} PNL^{-1}_{B, \odot}$$
In what follows, we adopt the distance of the Virgo cluster of D=15 Mpc
(\cite{Yasuda1997}).

The 36 ICPN candidates in our field excluding I-24 and I-33
imply a total associated luminosity of
$1.0\times10^{10}L_{B,\odot}$. However, some fraction of our candidates
may still turn out to be high redshift line emitters.
For this relatively bright sample selected from both {[OIII]} and
H$\alpha$ images, the fraction of high-z contaminants is probably
much less than the 25\% found by \citet{Arnaboldi2002a} for an
{[OIII]}-only selected sample. In order to derive a secure lower limit,
we reduce the associated luminosity to $7.5 \times 10^9 L_{B, \odot}$
based on the very conservative estimate.
The area surveyed by the Suprime-Cam is 0.196 deg$^2$. Thus,
the distance independent surface brightness of the diffuse light is
$\mu_B = 27.69$ mag arcsec$^{-2}$.

Next, we need to compare this with the light from the Virgo galaxies.
If the intracluster stellar population is closely related to the local
density
of galaxies, we should compare the diffuse light with the light of three
galaxies, M86($B_T$=9.83mag), M84(10.09), and NGC4388(11.76), which
dominates
the galaxy light in our field ($8.0 \times 10^{10}L_{B, \odot}$ in total).
This comparison gives an estimate of 9\% as the fraction of diffuse light
to galaxy light. On the other hand,
if the population is a large-scale phenomenon, we should compare
its surface brightness with the smoothed-out surface
brightness of Virgo galaxies.  In this case, we would have a substantially
larger fraction. It is, however, difficult to derive a reliable value with
our small sample. Since the inhomogeneous distribution seems to favor a
local
phenomenon, we take the value of $\sim$10\%.
Under the assumption that the intracluster stellar population has the
same mass-to-luminosity ratio as galaxies, $M_{\rm IC*}$ in equation (1) is
thus $\sim$10\% of $M_{\rm gal}$.

Now, we proceed to the estimate of $M_{\rm gas}$ and the total gravitating
mass, $M_{\rm total}$, to derive the baryonic fraction.
\citet{Schindler1999} made a detailed analysis of smoothed-out
distributions of galaxies and hot gas.
They decomposed the distribution into three subclusters identified around
M87, M86, and M49 (NGC4472).
They present the density profile of the main M87 subcluster
separately for hot gas, galaxies and total gravitating mass (their Fig.11b).
We read from their plot the relative contribution of respective components
{\it at the position of our field} as
$\rho_{total}:\rho_{gas}:\rho_{gal}$=60:8:3\footnote{We use density instead
of
mass hereafter.}

At the position of our field which is closer to M86, the contribution to
$\rho_{gal}$ from the M86 subcluster is 2.2 times that from the M87
subcluster
(their Fig. 6)\footnote{Contribution from the M49
subcluster is negligible since our field is far off the subcluster.}.
They found a compact X-ray halo around M86 subcluster. It is, however,
interpreted as interstellar matter stripped from M86 by ram pressure.
If this is the case, the gas is already counted as $\rho_{gal}$.
Accordingly, we do not increase $\rho_{gas}$ due to the M86 halo.
We adopt a constant mass-to-light ratio of $(M/L)_B$=10
({\it e.g.,} \cite{Gerhard2001}) for all the galaxies instead of
20 adopted by \citet{Schindler1999}. This halves the value of
$\rho_{gal}$ with respect to other components. Finally, we have
$\rho_{total}:\rho_{gas}:\rho_{gal}$=60:8:4.8(=(3+2.2$\times$3)/2).

With the contribution from the intracluster stellar population
estimated here, we have
$$\rho_{total}:\rho_{gas}:\rho_{gal}:\rho_{IC}=60:8:5:0.5$$
for the relative contribution, though $\rho_{IC}$ has a relatively large
uncertainty.
Therefore, if this field in the Virgo core is typical for the whole
cluster, about a half (=5.5/13.5) of the baryons  associated with the
cluster formed stars, and the fraction of baryonic matter is about
18\% (=13.5/73.5).
With the value $\Omega_b = 0.02$ for $h = 0.7$ (e.g., \cite{Fukugita98}),
we obtain $\Omega_{total}\simeq 0.11 (=0.02/0.18)$ for total
gravitating mass.

\vspace{3mm}
We thank the staff of Subaru Telescope for their support to
the commissioning of the Suprime-Cam and T.Hayashino, H.Tamura, and
Y.Matsuda for their help in measuring the characteristics of
the narrow band filters. This research is supported by the Grant-in-Aid
(13640231) from the Ministry of Education, Culture, Sports, Science
and Technology. O.G., K.C.F., and M.A. wish to thank the Swiss
National Foundation for research money and travel grants.
Data reduction was carried out at the computer system
operated by Astronomical Data Analysis Center
of the National Astronomical Observatory of Japan.


\clearpage

\setlength{\tabcolsep}{3pt}\footnotesize
\begin{longtable}{rrrrrrrrrrl}
  \caption{Candidates for ICPNe in the Virgo
  cluster.}\label{tab:table2}
\hline\hline
ID & \multicolumn{2}{c}{$\alpha~~~$(J2000)~~~$\delta$}&m{\small
(5007)}&{[OIII]}&
r(Kron)$^{1)}$&
H$\alpha$&$(V+R)$&{[OIII]-$(V+R)$}&{[OIII]-H$\alpha$}&Notes \\
\hline
\endhead 
\hline
\endfoot 
\hline
\multicolumn{11}{l}{\hbox to 0pt{\parbox{180mm}{\footnotesize
Notes: $^{1)}$ Kron half light radius is in units of pixels. $^{2)}$
confirmed spectroscopically. $^{3)}$ spectroscopy unsuccessful.}}}
\endlastfoot
\multicolumn{11}{c}{ICPNe Candidates in intracluster fields}\\
\hline
I-1 & 12:26:51.16 & 12:48:44.2  & 27.269&0.969&2.618& 1.995& -1.989&
2.958&-1.026& \\
I-2 & 12:26:50.95 & 12:49:48.4  & 27.193&0.893&2.440& 3.094&
5.000&-4.107&-2.200&  \\
I-3 & 12:26:36.70 & 12:54:12.0  & 26.700&0.400&2.801& 1.141& -0.075&
0.475&-0.741&  \\
I-4 & 12:26:36.47 & 12:54:21.3  & 26.891&0.591&2.187& 1.820& -1.759&
2.351&-1.229&  \\
I-5 & 12:26:34.35 & 12:54:56.6  & 27.269&0.969&2.292& 2.780&
5.000&-4.031&-1.811&  \\
I-6 & 12:26:30.50 & 12:51:26.5  & 27.100&0.800&2.320& 1.512&
5.000&-4.200&-0.711&  \\
I-7 & 12:26:29.96 & 12:54:09.0  & 26.982&0.682&3.426&-0.092& -2.177& 2.859&
0.774&  \\
I-8 & 12:26:20.71 & 12:38:26.2  & 26.389&0.089&2.099& 2.209& -1.343&
1.432&-2.120&  \\
I-9 & 12:26:19.57 & 12:34:10.0  & 26.677&0.377&2.781& 2.088& -2.634&
3.011&-1.710&  \\
I-10& 12:26:19.23 & 12:38:44.6  & 27.257&0.957&2.124& 1.472& -2.192&
3.149&-0.515&  \\
I-11& 12:25:46.03 & 12:54:14.8  & 27.207&0.907&2.137& 1.587&
5.000&-4.093&-0.680&  \\
I-12& 12:25:42.95 & 12:51:08.6  & 27.185&0.885&1.867& 1.753& -1.368&
2.253&-0.869&  \\
I-13& 12:25:39.80 & 12:43:41.6  & 26.926&0.626&2.173& 1.151&
5.000&-4.374&-0.525&  \\
I-14& 12:25:39.01 & 12:48:54.4  & 27.221&0.921&2.642& 1.956&
5.000&-4.079&-1.035&  \\
I-15& 12:25:38.03 & 12:55:30.0  & 27.022&0.722&2.595& 3.371&
5.000&-4.278&-2.649&  \\
I-16& 12:25:35.86 & 12:44:52.1  & 26.644&0.344&2.249& 2.291& -2.699&
3.043&-1.947& PN$^{2)}$\\
I-17& 12:25:35.36 & 12:51:46.2  & 26.919&0.619&3.259& 0.475& -2.199& 2.817&
0.143&  \\
I-18& 12:25:30.05 & 12:55:27.9  & 26.702&0.402&1.858& 2.237&
5.000&-4.598&-1.835&  \\
I-19& 12:25:28.24 & 12:53:27.0  & 26.840&0.540&2.194& 2.342&
5.000&-4.460&-1.802&  \\
I-20& 12:25:28.19 & 12:42:35.0  & 26.840&0.540&2.898& 2.180& -2.330&
2.870&-1.640&  \\
I-21& 12:25:27.99 & 12:54:11.1  & 27.138&0.838&2.047& 2.814&
5.000&-4.163&-1.977&  \\
I-22& 12:25:23.23 & 12:55:21.4  & 26.925&0.625&2.098& 2.412&
5.000&-4.376&-1.788&  \\
I-23& 12:25:21.61 & 12:54:25.4  & 27.092&0.792&2.356& 1.909& -1.580&
2.372&-1.117&  \\
I-24& 12:25:20.61 & 12:44:45.5  & 26.714&0.414&3.028& 2.530& -2.212&
2.626&-2.116& Ly$\alpha^{2)}$\\
I-25& 12:25:01.92 & 12:49:48.2  & 26.805&0.505&2.233& 1.968& -1.778&
2.283&-1.463&  \\
I-26& 12:24:54.95 & 12:47:25.4  & 26.691&0.391&3.451& 1.594& -0.855&
1.246&-1.204&  \\
I-27& 12:24:52.96 & 12:49:25.4  & 26.444&0.144&2.269& 1.840& -2.194&
2.338&-1.695&  \\
I-28& 12:24:48.21 & 12:46:53.1  & 27.272&0.972&2.508& 1.784& -1.381&
2.353&-0.812&  \\
I-29& 12:24:43.56 & 12:54:15.2  & 27.058&0.758&2.362& 1.863& -1.031&
1.789&-1.105&  \\
I-30& 12:24:39.76 & 12:53:57.0  & 27.137&0.837&2.071& 3.672&
5.000&-4.163&-2.835&  \\
I-31& 12:24:38.45 & 12:50:16.1  & 26.758&0.458&2.316& 0.890& -2.297&
2.755&-0.432&  \\
I-32& 12:24:36.88 & 12:52:27.2  & 27.187&0.887&2.303& 1.283&
5.000&-4.113&-0.397&  \\
\hline
\multicolumn{11}{c}{Over-luminous objects in M84}\\
\hline
I-33& 12:24:53.84 & 12:52:25.1 & 24.603&-1.697&2.144&-3.217& 5.000&-6.697&
1.521& $^{3)}$ \\
I-34& 12:25:06.70 & 12:54:52.6 & 25.795&-0.505&4.672& 1.104&
5.000&-5.505&-1.608&  \\
I-35& 12:25:03.11 & 12:54:24.2 & 25.977&-0.323&3.766&-0.241&-2.107&
1.784&-0.082&  \\
I-36& 12:25:01.25 & 12:51:19.8& 26.187&-0.113&3.143& 5.000&-2.794&
2.682&-5.113&  \\
I-37& 12:25:17.27 & 12:54:32.8& 26.283&-0.017&3.219& 1.972&-1.358&
1.341&-1.989&  \\
I-38& 12:25:07.36 & 12:55:02.9& 26.283&-0.017&4.800& 0.519&
5.000&-5.017&-0.536&  \\
\end{longtable}

pixels. $^{2)}$

\clearpage

\begin{figure}
\begin{center}
\end{center}
\caption{
Coadded {[OIII]} image of the $34'\times27'$ survey field. ICPNe
candidates are marked by the circles. The lower left area bounded by the
black line indicates the area excluded from our survey because of the dead
CCD chip. The margin of the field where S/N is low is also excluded.
Envelopes of bright galaxies are subtracted as much as possible for the
detection of PNe embedded there. }
\end{figure}

\begin{figure}
\begin{center}
\end{center}
\caption{
Finding chart of the 38 ICPNe candidates ([OIII] image).
An area $1'\times{1}'$ is shown. North is up, and east is to the left.}
\end{figure}

\end{document}